\documentclass[twocolumn]{aastex701}
\usepackage{graphicx}
\usepackage{subcaption}

\usepackage{amsmath}
\usepackage{float}
\usepackage{makecell}
\usepackage{array}

\usepackage{xcolor}
\usepackage{txfonts}
\usepackage[normalem]{ulem}

\begin{document}

\title{On the origin of the short-period Blue Straggler Stars in Collinder~261: a binary evolution approach}

\author[orcid=0009-0001-0083-2451,sname='Maite Echeveste']{M. Echeveste}
\affiliation{ Instituto de Astrof\'isica de La Plata, IALP (CONICET-UNLP), 1900 La Plata, Argentina}
\affiliation{INAF – Osservatorio Astrofisico di Arcetri, Via Largo E. Fermi 5, 50125 Firenze, Italy}
\email[show]{maite.echeveste@inaf.it}  

\author[0000-0003-4009-8316]{M.~J. Rain} 
\affiliation{European Southern Observatory, Alonso de C\'ordova 3107, Vitacura, Región Metropolitana, Chile}
\email[show]{mrainsep@eso.org}  

\author[0000-0002-0155-9434]{G. Carraro}
\affiliation{Dipartimento di Fisica e Astronomia, Universit\`{a} di Padova, Vicolo Osservatorio 3, I-35122, Padova, Italy}
\email{giovanni.carraro@unipd.it}  

\author[0009-0005-0973-470X]{O. G. Benvenuto}
\affiliation{ Instituto de Astrof\'isica de La Plata, IALP (CONICET-UNLP), 1900 La Plata, Argentina}
\affiliation{Facultad de Ciencias Astron\'omicas y Geof\'isicas, Universidad Nacional de La Plata (UNLP), La Plata, Argentina}
\email{obenvenuto@fcaglp.unlp.edu.ar}  

\author[0000-0002-6511-2463]{J. A. Panei}
\affiliation{ Instituto de Astrof\'isica de La Plata, IALP (CONICET-UNLP), 1900 La Plata, Argentina}
\affiliation{Facultad de Ciencias Astron\'omicas y Geof\'isicas, Universidad Nacional de La Plata (UNLP), La Plata, Argentina}
\email{panei@fcaglp.unlp.edu.ar}  

\author[0000-0002-8093-3024]{M. A. De Vito}
\affiliation{ Instituto de Astrof\'isica de La Plata, IALP (CONICET-UNLP), 1900 La Plata, Argentina}
\affiliation{Facultad de Ciencias Astron\'omicas y Geof\'isicas, Universidad Nacional de La Plata (UNLP), La Plata, Argentina}
\email{adevito@fcaglp.unlp.edu.ar}


\begin{abstract}

Building on the recent publication of a fresh catalog of blue straggler stars in Galactic open clusters using Gaia DR3, we conduct a study focusing on the blue straggler stars with short-period photometric variability found in the open cluster Collinder 261: nine eclipsing or ellipsoidal binaries and five SX Phe pulsator candidates
Our investigation delves into the potential of mass transfer in binary systems to account for the presence and properties of these stars. We first measure their photometric periods and classify the variables into eclipsing/ellipsoidal binaries, SX Phe pulsator candidates, and instrumental aliases or blends. Subsequently, we perform a set of simulations based on comprehensive calculations of the structure of both stars within the binary system, as well as their orbital evolution and mass transfer rates. To compare with observations, we solved the atmospheric structure of each stellar component and computed filter-weighted integrated fluxes in the Gaia DR3 passbands, obtaining a combined evolutionary sequence that accounts for the unresolved flux of both stars as seen photometrically. Our investigation includes the study of several parameters, such as the initial mass of both stars, the initial orbital period, and the efficiency of mass transfer. The aim is to identify potential progenitor candidates matching color-magnitude diagram observations at the same age as that determined for Collinder~261. We find that binary evolution serves as a viable mechanism for producing these eclipsing blue stragglers and that the efficiency of mass transfer plays a key role in accurately matching the fainter objects.

\end{abstract}

\keywords{\uat{Open star clusters}{1160} --- \uat{Blue straggler stars}{168} --- \uat{Stellar evolutionary models}{2046} --- \uat{Multiple star evolution}{2153}}

\section{Introduction} 
\label{sec:intro}

Blue-straggler stars (BSSs) are peculiar objects found in various stellar systems. Such peculiarity arises from their location in the optical color-magnitude diagram (CMD), being in a bluer and brighter region than their host stellar systems' main-sequence (MS) turn-off point. According to standard stellar evolution, stars in this region should have evolved off the MS long ago. To explain this unusual characteristic, the prevailing consensus is that BSSs were originally MS stars that acquired additional mass. This idea is supported by observations of BSSs in globular clusters that suggest that these stars are more massive than the stars in the MS turn-off point (\citealt{1997ApJ...489L..59S}, \citealt{1989ApJ...336..780N}). Three mechanisms are commonly proposed for the increase of the MS star mass: mass transfer (MT) in binary systems  (\citealt{McCrea_1964, Tian_2006, 2008MNRAS.387.1416C}), mergers in close binaries or hierarchical systems (\citealt{Andronov_2006, Knigge_2009, Perets_Fabrycky_2009, 2011MNRAS.410.2370L}), and direct collisions involving single, binary, or triple stars (\citealt{Hills_1976, 1989AJ.....98..217L, 1999ApJ...513..428S}).

In the MT BSSs' formation channel, the initially more massive star of the binary, referred to as the primary star, fills its Roche lobe. The star then transfers mass from its envelope to its companion, the secondary star, through the inner Lagrangian point $L_1$. This process is known as Roche lobe overflow (RLOF). Within this model, mass transfer through RLOF may be conservative (the total mass and angular momentum of the system are conserved) or non-conservative. As a result of the mass transfer process, the accreting star becomes a more massive MS star. This newly formed star occupies an upper place near the Zero Age Main Sequence (ZAMS) based on its new mass while continuing hydrogen burning in its core. Consequently, the MT process leads to the existence of an MS star with a longer lifespan compared to a normal star of the same mass.

BSSs formation via MT has been explored in previous studies. \citet{2004MNRAS.355.1182C} showed that Case~A and Case~B MT\footnote{Case~A: during core hydrogen burning of the donor (i.e., the star is still on the main sequence); Case~B: after hydrogen core exhaustion but before helium ignition of the donor (i.e., the star left the main sequence but has not reached the tip of the giant branch).} in low-mass binaries may produce BSSs with short (less than $\approx 40$~d) orbital periods. They also pointed out that the accreting star can be seen as bluer in the CMD as a result of abnormalities in its surface composition, such as unusual CNO abundances. They applied these results to the 4.2~d period BSS F190 in the $\approx 4~\rm Gyr$-old open cluster (OC) NGC~2682 (M67) and found that this BSS may be more likely formed through Case~A MT. Their simulations also show that F190 may still be in a slow stage of MT.\

\citet{Tian_2006} performed detailed stellar evolution calculations with the Eggleton code described in \citet{2000MNRAS.319..215H}, and designed a grid of models of close binaries that experience MT. In these models, they neglect the system's angular momentum loss due to magnetic braking and assume conservative MT. They used this grid to carry out a statistical study of M67. They concluded that other mechanisms, such as the merger of short-period binaries and wind accretion in wide binaries, must be operating in this cluster to account for the presence of its BSS population.

Among more recent works, we can mention those by \citet{2021ApJ...908....7S} and \citet{2023ApJ...944...89S}. In each of these articles, the authors conduct a detailed study of a specific BSS: WOCS 5379 and WOCS 4540, both located in the $\approx 6.5~\rm Gyr$-old OC NGC~188. These BSSs have a helium white dwarf companion in a wide orbit, suggesting an MT formation mechanism. Using the MESA code (\citealt{2011ApJS..192....3P}), the authors performed detailed calculations to identify the best models capable of explaining the main characteristics of these BSSs. In both studies, they allowed for the possibility of non-conservative MT. On the one hand, WOCS~5379 has an open orbit of 120~d and an eccentricity of 0.24, with the best-fitting model having an MT efficiency of 0.22 (the accreting star accretes 22\% of the mass transferred from the donor star). On the other hand, WOCS~4540 has an even wider orbit of 3030~d and an eccentricity of 0.36. The model that best fits this BSS has conservative MT by RLOF and non-conservative MT via stellar winds.

Using population synthesis models, \citet{Leiner_2021} studied the mechanism of MT in 16 OC BSS populations, showing that non-conservative MT plays an important role in producing realistic BBS masses.

More recently, \citet{Rain_2024} studied MT in binaries by performing a set of detailed calculations using an adaptation of the code described in \citet{2003MNRAS.342...50B}. They assumed conservative MT in all the models. They solved the atmosphere structure of both stars in the binary with an updated version of the code presented in \citet{Rohrmann2001MNRAS.323..699R} and calculated the evolutionary sequences in the CMD by summing both fluxes. They compared these simulations with the BSS population of twelve OCs, demonstrating excellent agreement between the bulk distribution of BSS and the composite evolutionary sequences in the CMD.

Here, we build upon the work of \citet{Rain_2024} by exploring the MT scenario to form the short-period (less than 2.1~d) BSSs found in Collinder~261. 

This OC is among the most ancient in the Milky Way, with published ages ranging from 5 to 11~Gyr depending on the adopted stellar models, reddening and distance \citep{Mazur_1995, Gozzoli_1996, 2002A&A...389..871D, 2006AJ....131.1544B, Netopil2016}, and most recent determinations between 6 and 7.5~Gyr \citep{Cantat-Gaudin_2020}. It has a near-solar metallicity, with a reddening {\it{E(B-V)}} between 0.25 and 0.34 mag, and is located at a distance of 2.2 to 2.9~kpc (\citealt{Mazur_1995}, \citealt{Gozzoli_1996}, \citealt{2002A&A...389..871D}, \citealt{2006AJ....131.1544B}). By analyzing the location of the BSSs in the CMD and inferring their orbital periods, we aim to investigate whether the MT formation mechanism can account for the presence and properties of these stars. 

The remainder of the paper is organized as follows. In Sect.~\ref{sec2} we outline the analysis of observational data and detail the methods used to determine periods. Sect.~\ref{sec3} summarizes the numerical method and physical assumptions made to perform the binary models and presents a comparison between them and the short-period BSSs in Collinder~261. Finally, in Sect.~\ref{sec4} we make a summary and elaborate our conclusions.

\section{Blue straggler population of Collinder~261}
\label{sec2}
\subsection{Isochrone fitting and blue straggler stars selection}

\begin{figure}
  \centering
  \includegraphics[width=0.5\textwidth]{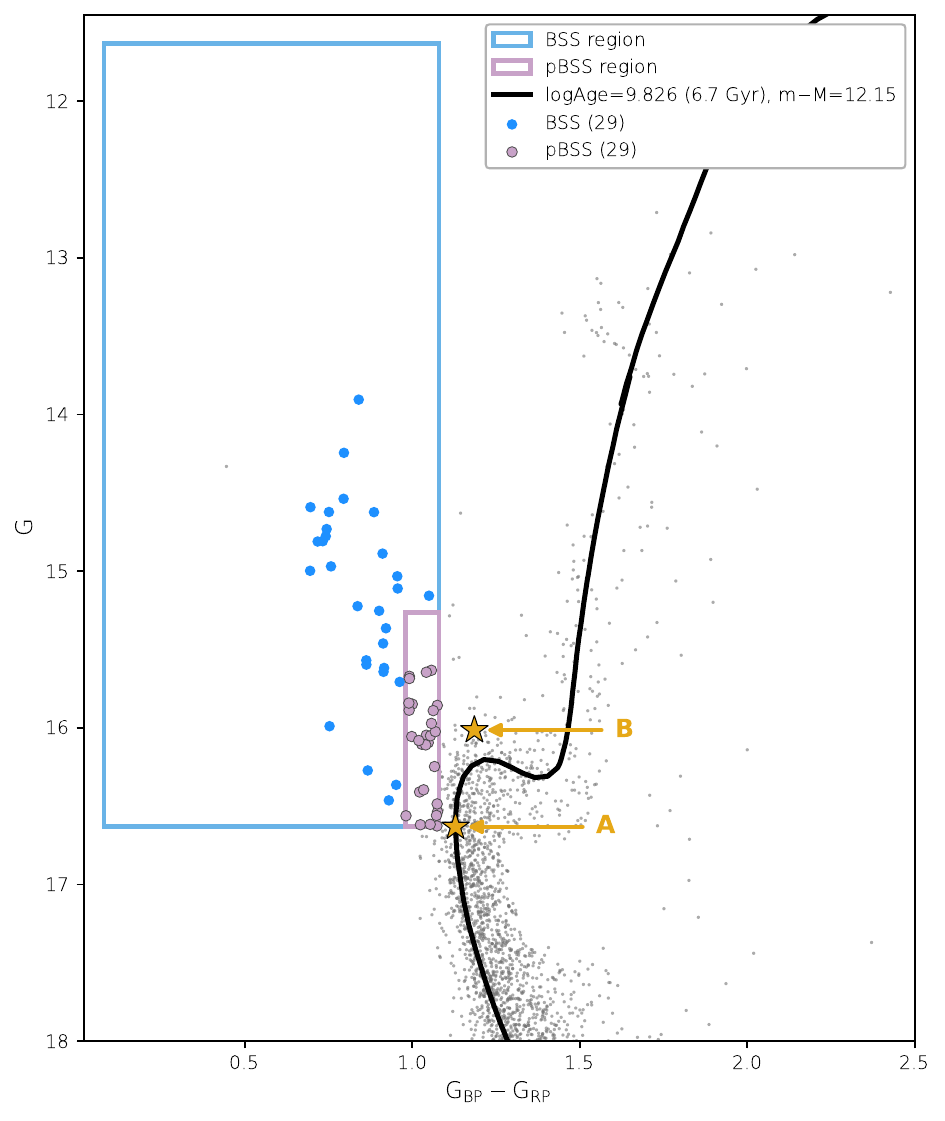}
    \caption{Metrically selected cluster members; stars deviating by more than 2$\sigma$ from the isochrone around the turnoff region were removed for clarity. The black line is the best-fitting PARSEC isochrone ($\log(\mathrm{age})=9.826$, i.e.\ 6.7\,Gyr, $m-M=12.15$). The golden arrows mark the two reference points of the \citet{Jadhav_2021} scheme, identified on the observed CMD: A, the main-sequence turnoff, and B, the brightest point of the main sequence prior to the onset of the subgiant branch. The light-blue rectangle shows the bona fide BSS region and the lilac box the pBSS region; the 29 BSS and 29 pBSS candidates are shown as blue and lilac filled circles, respectively (58 in total).}
    \label{ajuste}
\end{figure}

BSSs were identified using Gaia CMDs after selecting bona fide cluster members based on astrometric criteria; see \cite{Cantat-Gaudin_2018, Rain_2021_2} for more information. Here, a ridge line was constructed that traces the MS, subgiant branch (SGB), and red giant branch (RGB) by computing median colors and magnitudes in bins along the CMD, using the $G_{mag}$ versus $G_{\rm BP}-G_{\rm RP}$ plane. 

The PARSEC isochrones were calculated by adopting as starting point the values of age, distance, metallicity, and extinction given by \citet{Dias_2021}
($\log t = $ 9.8 $\pm$ 0.129, $d = $ 2.7~kpc, [Fe/H] $\sim$ 0, $A_V = $ 0.99 $\pm$v0.110). We then varied these parameters within their uncertainties to obtain the best match to the observed ridge line. The main-sequence turn-off (MSTO) was identified from the best-fitting isochrone as the point of infinite slope. The best-fitting isochrone has $\log t = 9.826$ (6.7~Gyr), $E(B-V) = 0.98$  and a true distance modulus $(m-M)_0 = 12.15$ ($d = 2.69$~kpc). This age lies within the range of published values for Collinder~261, between the 6~Gyr of \citet{2006AJ....131.1544B} and the 7--8~Gyr of \citet{Gozzoli_1996}, and is consistent within the uncertainties with the homogenised value of \citet{Netopil2016} ($\log t = 9.86 \pm 0.17$). Older isochrones ($\gtrsim 7.5$~Gyr) produce a turn-off that is too faint and red to reproduce the observed ridge line for the adopted distance and reddening. Since the binary models of Section~\ref{sec3} are compared to this same Gaia DR3 CMD, we adopt 6.7~Gyr as the cluster age throughout the paper. 

Following the scheme introduced by \cite{Jadhav_2021}, two reference points were defined in the CMD: see Figure \ref{ajuste}. Point A corresponds to the MSTO and is identified as the bluest point along the main sequence, and point B corresponds to the brightest point of the main sequence prior to the onset of the SGB. The region of bona fide BSSs was defined as a rectangular region extending approximately 1 mag towards bluer colors and up to 5 mag towards brighter magnitudes relative to point A. Stars located within this region are classified as BSSs. In addition, a region of probable blue straggler stars (pBSSs) was defined to account for stars located very close to the MSTO, where photometric scatter and unresolved binaries can affect the classification. This pBSS region extends approximately 0.1 mag in color and has an upper magnitude limit set at 0.75 mag brighter than point B, corresponding to the tip of the main sequence. While this region contains genuine blue stragglers, it is also expected to include some main-sequence or MS-tip stars, as mentioned in \cite{Rain_2021_2, Rain_2024}. We identified a total of 58 blue straggler candidates within Collinder~261.

\subsection{TESS Insights into the variability of blue straggler stars}

\begin{table}
	\centering 
	\caption{Photometric periods of the 26 BSS candidates with short-period photometric signals in Collinder~261, with their variability classification. For the eclipsing/ellipsoidal binaries, the adopted orbital period is $P_{\rm orb} = 2\,P_{\rm LS}$. Stars flagged as ``Blend'' share the signal of the neighbouring variable given in parentheses. ``Instrumental alias'' marks the four stars sharing the $P \approx 0.402$~d systematic signal. For the tentative long-period variables the period is $P_{\rm LS}$ as measured in a single TESS sector.}
	\label{tab:short_periods_Cr261}
    \setlength{\tabcolsep}{3pt}
    \begin{tabular}{clll} 
		\hline
		\emph{Gaia} DR3~ID  &  TIC~ID  & Period & Type\\ 
		 &  & [d] \\
		\hline
5856515601255190272 & 327546480 & 2.110 & Eclipsing \\  
5856515665648722688 & 327546384 & 1.420 & Eclipsing \\  
5856517010004497792 & 327766459 & 1.048 & Eclipsing \\  
5856527386646253312 & 327546366 & 1.318 & Eclipsing \\  
5856527455365735680 & 327546290 & 0.402 & Instrumental alias\\
5856527455365740032 & 327546276 & 0.214 & Blend (TIC 327546246) \\
5856527455365743488 & 327546246 & 0.214 & SX Phe pulsator\\
5856527524085202048 & 327546373 & 0.166 & Blend (TIC 327546308)\\
5856527524085206656 & 327546308 & 0.166 & SX Phe pulsator\\
5856527558444960128 & 327546189 & 0.401 & Instrumental alias \\
5856527558444963200 & 327546164 & 0.402 & Instrumental alias\\
5856527661524172800 & 327546215 & 0.402 & Instrumental alias\\
5856527730243676800 & 327545961 & 1.092 & Eclipsing\\ 
5856527764603421184 & 327545942 & 0.200 & SX Phe pulsator\\
5856528383078700288 & 327546013 & 1.090 & Eclipsing\\
5856528486157941376 & 327545791 & 0.982 & Blend (TIC 327545804) \\
5856528550550957184 & 327545804 & 0.982 & Eclipsing\\
5856528623597294720 & 327545871 & 0.230 & SX Phe pulsator\\
5856528761035882496 & 327545566 & 0.518 &Eclipsing\\
5856528962867671936 & 327546031 & 0.200 & SX Phe pulsator\\
5856432828647768960 & 327153496 & 1.128 & Eclipsing\\
\hline
5856528348719366528 & 327546103 & 3.396 & Long-period (tentative) \\
5856529379511194240 & 327767012 & 3.635 & Long-period (tentative) \\
5856528726676132096 & 327545609 & 3.663 & Long-period (tentative) \\
5856528348719355648 & 327546158 & 3.704 & Long-period (tentative) \\
5856511271928032256 & 327765836 & 4.660 & Long-period (tentative) \\

		\hline
	\end{tabular}
\end{table}

Transiting Exoplanet Satellite Survey (TESS)'s continuous monitoring allows for the detection of periodic variability, eclipses, and other transient phenomena. TESS data enable us to observe these variations with a level of detail that was previously unattainable, helping to distinguish between different formation scenarios of BSSs such as MT in binary systems or stellar collisions. Both, the 2-min and 30-min cadence are particularly well-suited for discovering short-period binaries because they provide dense temporal coverage, ensuring that rapid changes in brightness associated with eclipses are captured with high precision. While the 2-min cadence is focused on pre-selected targets, the 30-min cadence allows for the discovery of binaries that might not have been previously identified or targeted.
Over a 27-day observation period, even with a 30-min cadence, TESS can capture numerous data points, which is enough to reveal periodic variations for short-period binaries. Although it might miss the finer details of rapid changes within a very short period binary (e.g., a binary with an hour-long period), it can still clearly show the periodicity and overall light curve shape for binaries with periods of several hours or longer.

Our analysis was performed on the light curves (LC) of the project “a PSFs-based approach to TESS High-quality data Of Stellar clusters” (PATHOS, \citealt{Nardiello_2019}), born originally to exploit TESS data to extract high-precision photometry for members of stellar clusters and associations, adopting an innovative approach based on the use of empirical Point Spread Functions (PSFs) and neighbor subtraction. PATHOS  LCs are publicly available and released on the Mikulski Archive for Space
Telescopes (MAST) as High-Level Science Product (HLSP). 

For all our BSS candidates, we searched the TESS light curves. To find the periodicity, we used the classical Lomb-Scargle periodogram (named for \citealt{Lomb_1976} and \citealt{Scargle_1982}), which is a classic method for finding periodicity in irregularly sampled data. The periodogram itself occupies a unique niche: it is motivated by Fourier analysis, but it can also be viewed as a least squares method. In this context, the \texttt{gatspy}\footnote{\url{http://www.astroml.org/gatspy/}} package \citep{VanderPlas_2015, gastpy_2015} is a public python-friendly algorithm that provides three different main implementations of the classic Lomb-Scargle periodogram; LombScargle, LombScargleFast and LombScargleAstroML. In this work, we have used the LombScargleFast algorithm, which implements an efficient pure-Python version of the Fast Fourier Transform-based algorithm from \cite{Press_1989}. The latter is much faster than either of the remaining implementations of \texttt{gatspy}, especially as the number of data points and frequencies increases. It is limited, however, to either a simple pre-centered model or a floating mean model, and the frequencies must be computed on a regular grid.

\begin{figure*} 
	\includegraphics[width=2\columnwidth]{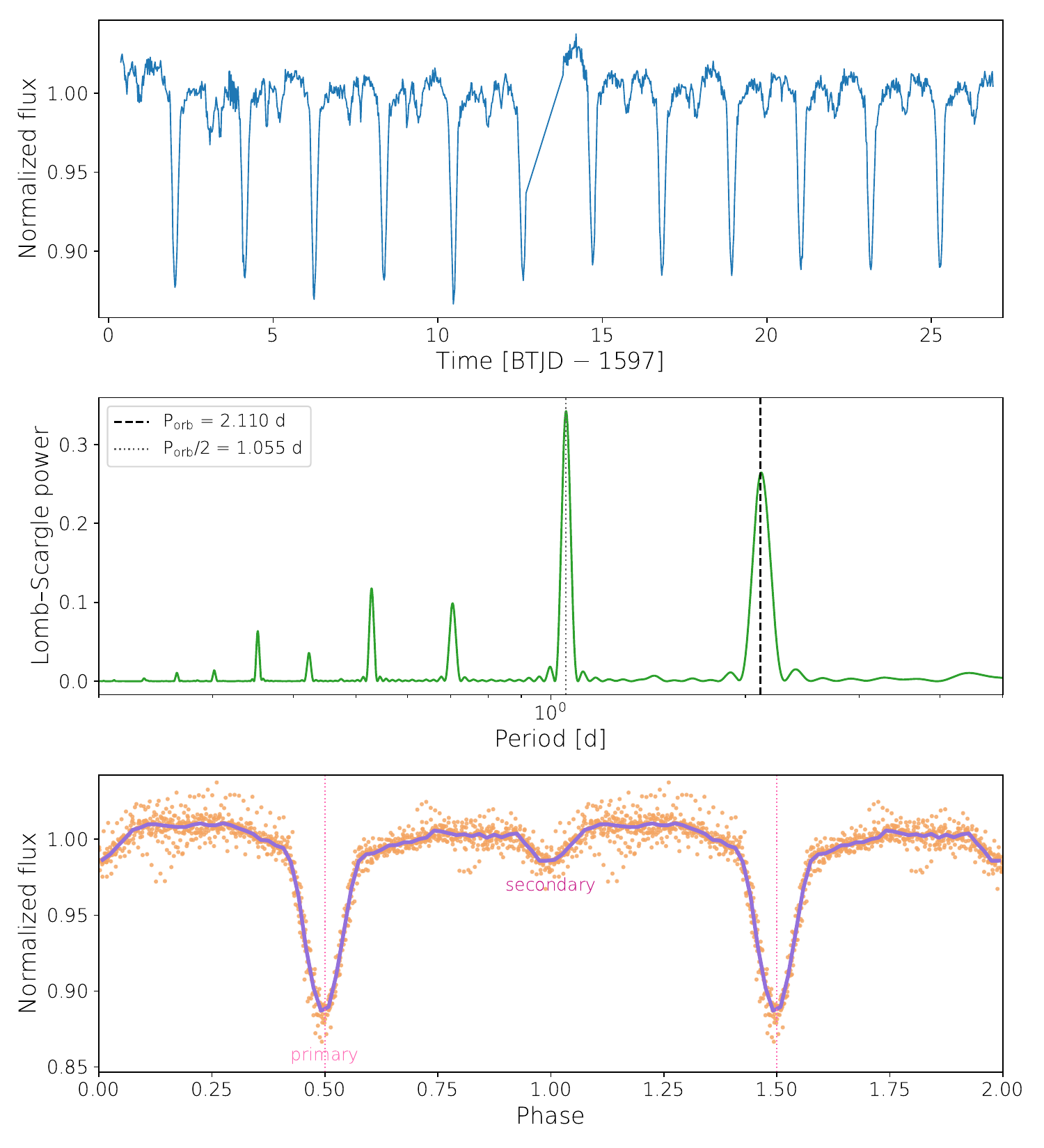}
    \caption{TESS/PATHOS photometry of TIC~327546480 (Gaia DR3 5856515601255190272), an eclipsing binary in the field of Collinder~261, from Sector~11. Top: normalized PATHOS PSF light curve. Middle: Lomb--Scargle periodogram; the dashed line marks the adopted orbital period $P_{\rm orb} = $ 2.110~d and the dotted line its first harmonic $P_{\rm orb}/2 = $ 1.055~d. Bottom: light curve phase-folded on $P_{\rm orb}$, with the binned median overplotted (purple). The unequal eclipse depths (primary $\approx 11\%$, secondary $\approx 1\%$) confirm the orbital period as twice the dominant photometric signal and indicate components of distinct effective temperature.}
\label{fig:tic327546480_aperture}.  
    \label{fig:LC_example}
\end{figure*}

Since extreme outliers can often complicate the analysis of TESS LC's, after we downloaded the data, we "cleaned" each LC following a similar procedure to the one reported by \cite{Nardiello_2020}; we excluded all the points whose: i) quality flag is higher than Zero \texttt{DQUALITY} $> 0$, ii) those with values above 2.5~$\sigma_\mathrm{flux}$ the median flux, and iii) with values higher than 3.5~$\sigma_\mathrm{sky}$ the median value of the sky. 
Additionally, when considering the periodogram from the Fourier perspective, it is useful to normalize the periodogram so that, in the special case of equally spaced data, it recovers the standard Fourier power spectrum. On this line, to define the continuum, we binned our data in 10~bins of $\sim$2.5 days each and we estimated the highest points in each bin.  Then, we interpolated these points with splines and divided each point of the LC's by this interpolation.  In Figure \ref{fig:LC_example}, an example is shown.

In total, 26 of the 58 BSS candidates show significant short-period photometric signals. The Appendix allows us to separate them into nine eclipsing or ellipsoidal binaries with orbital periods $P_{\rm orb} = 2\,P_{\rm LS}$ between 0.52 and 2.11~d, five SX~Phoenicis pulsator candidates with pulsation periods of 0.16--0.23~d, three stars whose signal is a blend of a neighbouring variable BSS, four stars sharing an instrumental alias at $P \approx 0.402$~d, and five tentative long-period variables. The periods and types, together with the Gaia and TESS IDs, are reported in Table~\ref{tab:short_periods_Cr261}. The resulting fraction of confirmed short-period binaries among the BSS candidates is therefore $\sim$15.5\% (9/58).
These detections are consistent with previous studies of the variability of this group. \citet{Mazur_1995} discovered 45 short-period eclipsing binaries in the field of Collinder\,261, 28 of them of W~UMa type, with the detached and semidetached systems concentrating at periods of order one day, which is precisely the period range we recover for our nine eclipsing systems. 
Three pairs of BSS candidates separated by less than $\sim$1 TESS pixel (13--23\arcsec) show identical periods ($P_{\rm orb} = 0.982$~d, $P = 0.214$~d and $P = 0.166$~d). In each pair, the PATHOS light curves of the two stars have the same period, phase, and absolute amplitude in the larger apertures, as expected for a single variable source, while in the smallest aperture (1~px radius), the amplitude is 1.4--1.8 times larger when centered on the brighter star. We therefore assign the variability to TIC~327545804, TIC~327546246, and TIC~327546308, and flag their neighbors (TIC~327545791, TIC~327546276, and TIC~327546373) as blends. The amplitude does not increase between the two largest apertures, which excludes the bright ($G \approx 12$) stars located within 3 pixels as the source of the signal.

At face value, our short-period binary fraction is higher than the $\sim$10\% frequency of \emph{spectroscopic} BSS binaries found in NGC\,188 and M\,67 \citep{Mathieu_2009, Boffin_2015}. However, the two techniques probe nearly disjoint period regimes. The RV-detected BSS binaries in those clusters have typical periods of $\sim$$10^3$\,d, while sub-day contact and near-contact systems are challenging for RV surveys: their components are tidally spun up, producing broad, blended spectral lines, and their photometric variability itself complicates precise RV measurements. Conversely, TESS photometry over a single sector is only sensitive to periods of up to a few days. The two fractions are therefore complementary rather than contradictory, and their sum suggests that a large fraction of BSSs in old open clusters reside in binaries, as expected in the MT formation scenario. We also note that some of our systems could be members of hierarchical triples, in which case the photometric period would trace the inner binary; the present data do not allow us to test this possibility. Finally, genuine pulsators are explicitly identified and excluded from the binary fraction: the five SX~Phe candidates show the single-minimum, asymmetric folded curves and sub-6-hour periods characteristic of this class, unlike the double-minima morphology of the eclipsing systems (Appendix~\ref{app:lightcurves}).

\section{Binary models}
\label{sec3}

\subsection{Numerical method and physical assumptions}

To account for the presence and properties of the short-period eclipsing BSSs in Collinder~261, we perform a set of calculations using a variation of the code for stellar binary evolution presented in \cite{2003MNRAS.342...50B}. Briefly, this code solves the equations of stellar structure together with the orbital evolution, considering the system is circularized and synchronized, and the stars exchange mass through RLOF. First, the evolution of the stellar structure of the primary star is calculated from the ZAMS as if it were isolated. Once this star fills its Roche lobe, it transfers mass through the inner Lagrangian point $L_1$, and the code calculates the structure of the star self-consistently with the mass transfer rate (MTR) (i.e., the MTR is included as an extra variable in the Henyey scheme). The code saves the amount of matter lost by the star in each step, which will be used for calculating the amount of matter accreted by the secondary star. Here, we introduce the parameter $\beta$, which regulates the efficiency of the MT process. The relation between the mass lost by the primary and the mass accreted by the secondary star is then 

\begin{equation}
  \dot{M}_{2}= -\beta \dot{M}_{1}, 
  \label{eq:beta}
\end{equation}

\noindent where $0 \leq \beta \leq 1$. Here, suffix 1 (2) refers to the primary (secondary) star. The angular momentum lost by the system when the MT is non-conservative (i.e. $\beta <1$) involves another free parameter commonly referred to as $\alpha$ (\citealt{1982ApJ...254..616R, 1983ApJ...275..713R}). This parameter accounts for the angular momentum carried away by matter lost from the system in units of the specific angular momentum of the accretor.

The calculations stop at a chosen age, which for our purposes corresponds to the age of Collinder~261. Next, we calculate the evolution of the secondary star's stellar structure from the ZAMS, assuming that the system is detached. Once the age at which the primary star begins transferring mass is reached, the structure of the secondary is computed as if it were accreting mass, following Eq. \ref{eq:beta}. From then on, four scenarios are possible: (1) MT from the primary star stops before the cluster age and the binary becomes detached, (2) the secondary star keeps accreting mass while staying inside its Roche Lobe, resulting in a semidetached binary until the cluster age is reached, (3) the secondary star fills its Roche Lobe while the primary star is transferring mass, leading to contact between the binary components before the cluster age is attained, and (4) MT starts in the opposite direction, from the secondary star to the primary star (only possible if the MT from the primary stops at some point before reaching the cluster age).

We consider non-rotating stars, a mixing length parameter value tuned to reproduce the present Sun, moderate overshooting, semiconvection as in \citet{1983A&A...126..207L}, and thermohaline mixing following \citet{2013A&A...553A...1M}. The chemical evolution of the models was solved considering non-instantaneous mixing \citep{1985A&A...145..179L}. To compare with BSSs data from Collinder~261, we assume all stars in the simulations have solar metallicity. As we deal with short-period binaries in which tidal interactions are effective, we assume that all systems are circularized at all ages. We take into account the angular momentum loss due to magnetic braking \citep{verbunt1981A&A...100L...7V,1983ApJ...275..713R}. 

To compare the evolutionary tracks with the location of the observed BSSs in the CMD, we need to transform the effective temperature and bolometric luminosity into color and magnitude with Gaia DR3 filters. This requires first solving the atmospheric structure of each star in the binary and computing the filter-weighted integrated fluxes for each model using DR3-passbands. We solve the atmospheric structure under the assumption of local thermodynamic equilibrium. The calculations were performed using an updated version of the code described by \cite{Rohrmann2001MNRAS.323..699R}, which assumes hydrostatic and radiative-convective equilibrium. For models with cold atmospheres, convective transport was handled using the mixing-length approximation. After obtaining the evolutionary sequence of each star in the CMD, we calculate the magnitude of the binary by summing the flux of each star. This results in a combined evolutionary sequence that is used to compare with the BSS data. More details can be found in \cite{Rain_2024}.

\subsection{Results: Model-Observed BSSs comparison}

\begin{table}
    \centering
    \begin{tabular}{c c c c c} 
     $M_1$  & $M_2$ & $P_{\rm orb,i}$ & STATUS & $P_{\rm orb,T_C}$ \\
     $[M_{\odot}]$ & $[M_{\odot}]$ & $[\rm days]$ &  & $[\rm days]$ \\
    \hline
   1.11  & 0.74 & 0.65 & no MT        &      \\
   ''    & ''   & 0.60 & contact 6.59 & 0.40  \\ 
   1.14  & 0.76 & 0.85 & no MT        &      \\
   ''    & ''   & 0.80 & BSS          & 0.58 \\
   ''    & ''   & 0.75 & BSS          & 0.51 \\
   ''    & ''   & 0.70 & BSS          & 0.42 \\
   ''    & ''   & 0.65 & contact 6.13 & 0.42 \\
   1.17  & 0.78 & 1.10 & no convergence  \\
   ''    & ''   & 1.05 & BSS          & 1.80 \\
   ''    & ''   & 1.00 & BSS          & 1.53 \\
   ''    & ''   & 0.95 & BSS          & 1.27 \\
   ''    & ''   & 0.90 & BSS          & 0.96 \\
   ''    & ''   & 0.85 & BSS          & 0.73 \\
   ''    & ''   & 0.80 & BSS          & 0.54 \\
   ''    & ''   & 0.75 & BSS          & 0.40 \\   
   ''    & ''   & 0.70 & contact 5.69 & 0.44 \\
   1.20  & 0.80 & 1.00 & BSS          & 2.49 \\
   ''    & ''   & 0.95 & BSS          & 1.56 \\
   ''    & ''   & 0.90 & BSS          & 1.12 \\
   ''    & ''   & 0.85 & BSS          & 0.70 \\
   ''    & ''   & 0.80 & contact 6.43 & 0.46 \\
   1.23  & 0.82 & 1.00 & BSS          & 2.76 \\
   ''    & ''   & 0.95 & contact 6.53 & 1.92 \\
   
    \hline
    \end{tabular}
    \caption{List of initial parameters of the binary models ($M_1, M_2, P_{\rm orb,i}$) and the orbital period at the age $T_C$ ($P_{\rm orb, T_C}$) for the first set of simulations ($q=1.5$). The column STATUS indicates whether each binary becomes a BSS ({\it{BSS}}), enters contact ({\it{contact}}, followed by the age at which contact is reached in Gyr), or MT has not started within the age $T_c$ ({\it{no MT}}). For the binaries that reach contact, the orbital period in the column $P_{\rm orb,T_C}$ corresponds to the period at the moment of contact. All models were evolved with MT efficiency $\beta=1.0$.}
    \label{q150beta100}
\end{table}



Four fundamental quantities are necessary to characterize a binary system: the masses of the two stars, the orbital period, and the MT efficiency. These four quantities define the initial free parameter space of our simulations. We do not explore the effects of varying the parameter $\alpha$ in the MT process, but we used a fixed value $\alpha=1$. 
Here, we will look for possible progenitors of the nine eclipsing BSSs with orbital periods between 0.52 and 2.11 days listed in Table~\ref{tab:short_periods_Cr261}. We define BSS progenitors as binaries that, at the cluster age, have combined evolutionary sequences that place them within the region populated by the observed eclipsing BSSs in the CMD, and whose orbital periods fall within the observed range.

In our simulations, we look for systems that, by the age of the cluster $T_C$, have undergone MT. At this point, we separate two possible situations: the binary is semidetached or detached at age $T_C$, or the system reaches a contact configuration before age $T_C$. Our code does not follow the evolution after contact, and although the calculations stop at the onset of this configuration, this does not necessarily imply immediate coalescence. Low-mass contact binaries may survive for timescales of the order $10^8 - 10^9$ years before merging 
\citep{rasio1995ApJ...444L..41R,Qian2003Are,Tutukov2004Evolutionary,stepien2004IAUS..219..967S,liu2018MNRAS.474.5199L,Jiang2026Merger}. Contact binaries could therefore serve as possible progenitors of our eclipsing BSSs if they survive without merging until the age of the cluster. In fact, W~UMa binaries have been found within the BSS region of open and globular clusters \citep{Kaluzny1987ApJ...314..585K,Mateo1990Blue,Rucinski2000AJ....120..319R,Li2013Two,Dalessandro_2013}, and merging contact binaries have been proposed to populate the BSS sequence in the CMD of old clusters \citep{Stepien_2015}. 

As a first step, we adopt a fixed mass ratio of $q = M_1/M_2= 1.5$ and use a mass step of $0.03 M_{\odot}$ to explore the limits on the initial stellar masses that satisfy the conditions for possible progenitors. The value of the initial orbital period, $P_{\rm orb,i}$, must be small enough to allow the MT process to start before the age of the cluster is reached. We used a step of 0.05 days in the initial orbital period. Each combination of mass and orbital period is calculated for MT efficiency $\beta = 1.0$. The initial primary mass is constrained by inspecting the position of the combined evolutionary sequence in the CMD at age $T_C$: masses below a minimum value produce tracks that fall below the BSS region, while masses above a maximum value place the combined sequence above the brightest observed BSSs. Only those models whose tracks fall within the region populated by the observed BSSs in the CMD are retained as viable progenitor candidates. For each initial mass, we find that $P_{\rm orb,i}$ is limited by a minimum value below which the system enters contact and a maximum value above which there is no MT within the age $T_C$, or the orbital period at age $T_C$, $P_{\rm orb,T_C}$, is too large to be considered compatible with the short periods observed in our eclipsing BSS sample. Table \ref{q150beta100} shows the initial parameters of the simulations ($M_1$, $M_2$, $P_{\rm orb,i}$) indicating whether each binary becomes a BSS progenitor candidate (labeled as {\it{BSS}}), enters contact before age $T_C$ (labeled as {\it{contact}} followed by the age in Gyr at which contact is reached), or MT has not started within the age $T_C$ (labeled as {\it{no MT}}). The column $P_{\rm orb,T_C}$ reports the orbital period at age $T_C$ for the systems labeled as BSSs, and the orbital period at the onset of contact for those labeled as contact. Notice that there is one system labeled as {\it{no convergence}} for which the mass transfer rate gets too large ($\sim 10^{-3.2} M_{\odot}/yr$) for the code to converge. This system, together with the other two flagged in Tables \ref{q200beta100} and \ref{q150beta050}, starts MT between 1 and 1.5 Gyrs later than the more massive systems at the same (or even shorter) initial orbital period. 
Because of this delay, the donor star undergoes RLOF, at a more advanced evolutionary stage, with a deeper convective envelope. Since the response of the donor to mass loss is 
strongly affected by the type of energy transport at the envelope, in these conditions even a modest mass loss leads to the large MT rate values seen in these three systems.

\begin{figure*}
    \centering
    \includegraphics[width=\linewidth]{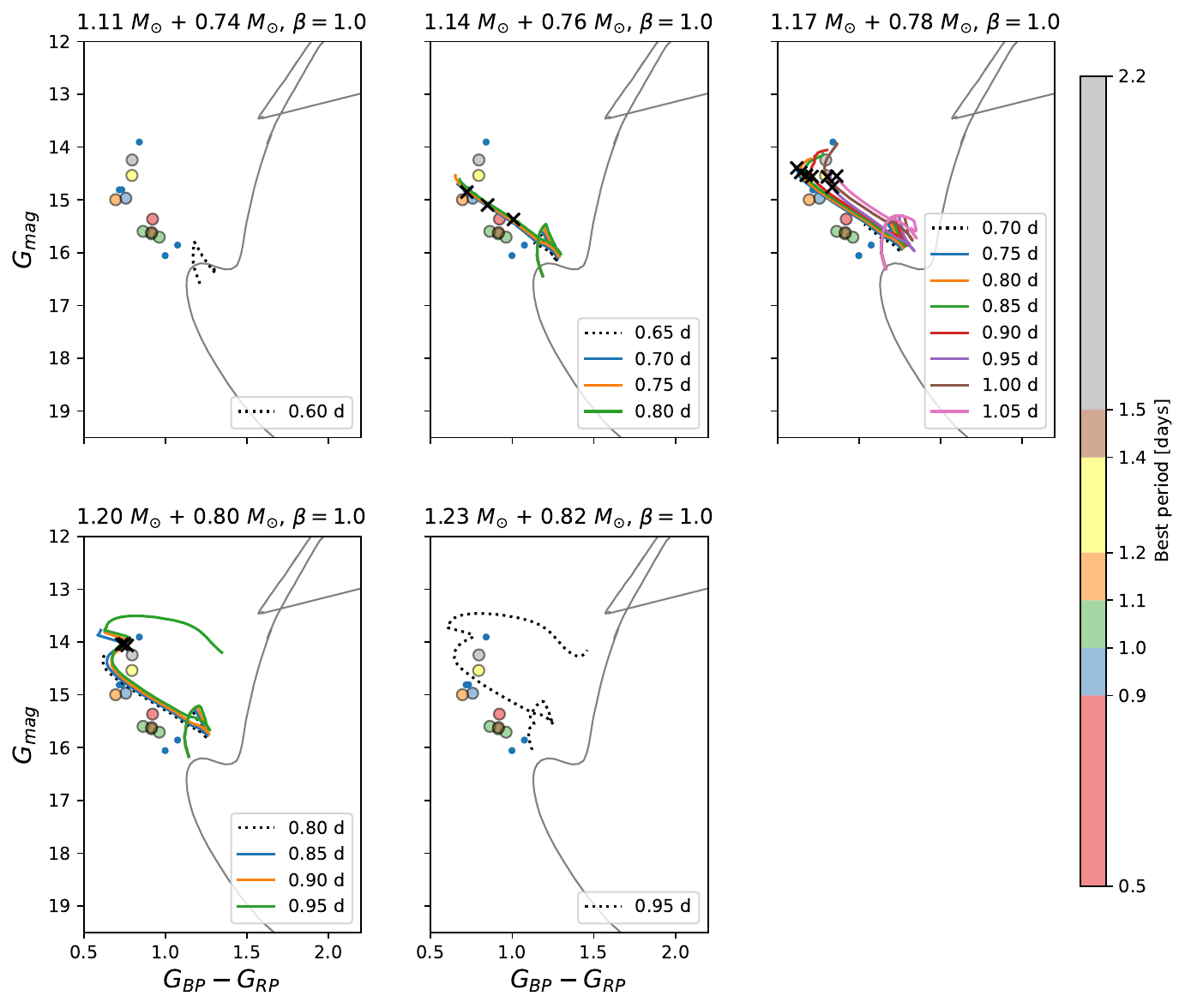}
    \caption{Combined evolutionary sequence of the BSS progenitors listed in Table~\ref{q150beta100} in the CMD. Each panel corresponds to a family of models with the same fixed initial masses, with tracks labeled according to the value of $P_{\rm orb,i}$. Dotted curves correspond to tracks that reach contact a few hundred million years before age $T_C$, and a black cross marks the point at which each model reaches $T_C$. The positions of the observed eclipsing BSSs are shown as circles colored by their best observed orbital period (Table~\ref{tab:short_periods_Cr261}), as indicated by the vertical color bar. Blue dots show tshow the positions of the five SX~Phe pulsators. The 6.7~Gyr isochrone is also shown for reference. }
    \label{fig:q150beta100}
\end{figure*}

Figure \ref{fig:q150beta100} shows the combined evolutionary sequence of the models listed in Table \ref{q150beta100} and the isochrone as a reference. We only show those models that represent good progenitors, i.e. with $P_{\rm orb,T_C} \leq 2.11$ days. Each panel corresponds to a set of simulations with the same initial masses, as read in the title. In each panel, different tracks correspond to different values of $P_{\rm orb,i}$. A black cross marks the point at which each model reaches age $T_C$. This marker is absent in the dotted curves, which correspond to systems that reach contact before age $T_C$. The positions of the observed eclipsing BSSs are shown with circles colored by orbital period. Most models reach age $T_C$ at magnitudes lower than $\sim 15$, since the accretor has had sufficient time to gain mass through MT. Models that reach fainter magnitudes fall into two categories: those with long $P_{\rm orb,i}$, and contact binaries. In both cases, the accretor had less time to accrete mass, resulting in a fainter combined magnitude. Notably, the five SX~Phe pulsator candidates also occupy this same region of the CMD, coinciding with the position reached by our models at age $T_C$; their possible connection to binary MT is discussed in Section~\ref{sec4}.

\begin{table}
    \centering
    \begin{tabular}{c c c c c} 
     $M_1$  & $M_2$ & $P_{\rm orb,i}$ & STATUS & $P_{\rm orb,T_C}$ \\
     $[M_{\odot}]$ & $[M_{\odot}]$ & $[\rm days]$ &  & $[\rm days]$ \\
    \hline
   1.11  & 0.555 & 0.60 & no MT        &      \\
   ''    & ''   & 0.55 & contact 6.30 & 0.33  \\ 
   1.14  & 0.57 & 0.80 & no MT        &      \\
   ''    & ''   & 0.75 & BSS          & 0.53 \\
   ''    & ''   & 0.70 & BSS          & 0.44 \\
   ''    & ''   & 0.65 & contact 6.39 & 0.38 \\
   1.17  & 0.585& 1.00 & no convergence  \\
   ''    & ''   & 0.95 & BSS          & 1.43 \\
   ''    & ''   & 0.90 & BSS          & 1.07 \\
   ''    & ''   & 0.85 & BSS          & 0.78 \\
   ''    & ''   & 0.80 & BSS          & 0.57 \\
   ''    & ''   & 0.75 & BSS          & 0.41 \\   
   ''    & ''   & 0.70 & contact 6.64 & 0.31 \\
   1.20  & 0.60 & 1.00 & BSS          & 2.91 \\
   ''    & ''   & 0.95 & BSS          & 1.60 \\
   ''    & ''   & 0.90 & BSS          & 1.40 \\
   ''    & ''   & 0.85 & BSS          & 0.76 \\
   ''    & ''   & 0.80 & BSS          & 0.45 \\
   ''    & ''   & 0.75 & contact 6.40 & 0.34 \\
   1.23  & 0.615& 1.00 & BSS          & 2.63 \\
   ''    & ''   & 0.95 & BSS          & 1.86 \\
   ''    & ''   & 0.90 & BSS          & 1.65 \\
   ''    & ''   & 0.85 & BSS          & 0.63 \\
   ''    & ''   & 0.80 & contact 6.20 & 0.40 \\   
   
    \hline
    \end{tabular}
    \caption{Same as Table \ref{q150beta100} but for those models in the second set of simulations ($q=2.0$, $\beta=1.0$).}
    \label{q200beta100}
\end{table}

\begin{figure*}
    \centering
    \includegraphics[width=\linewidth]{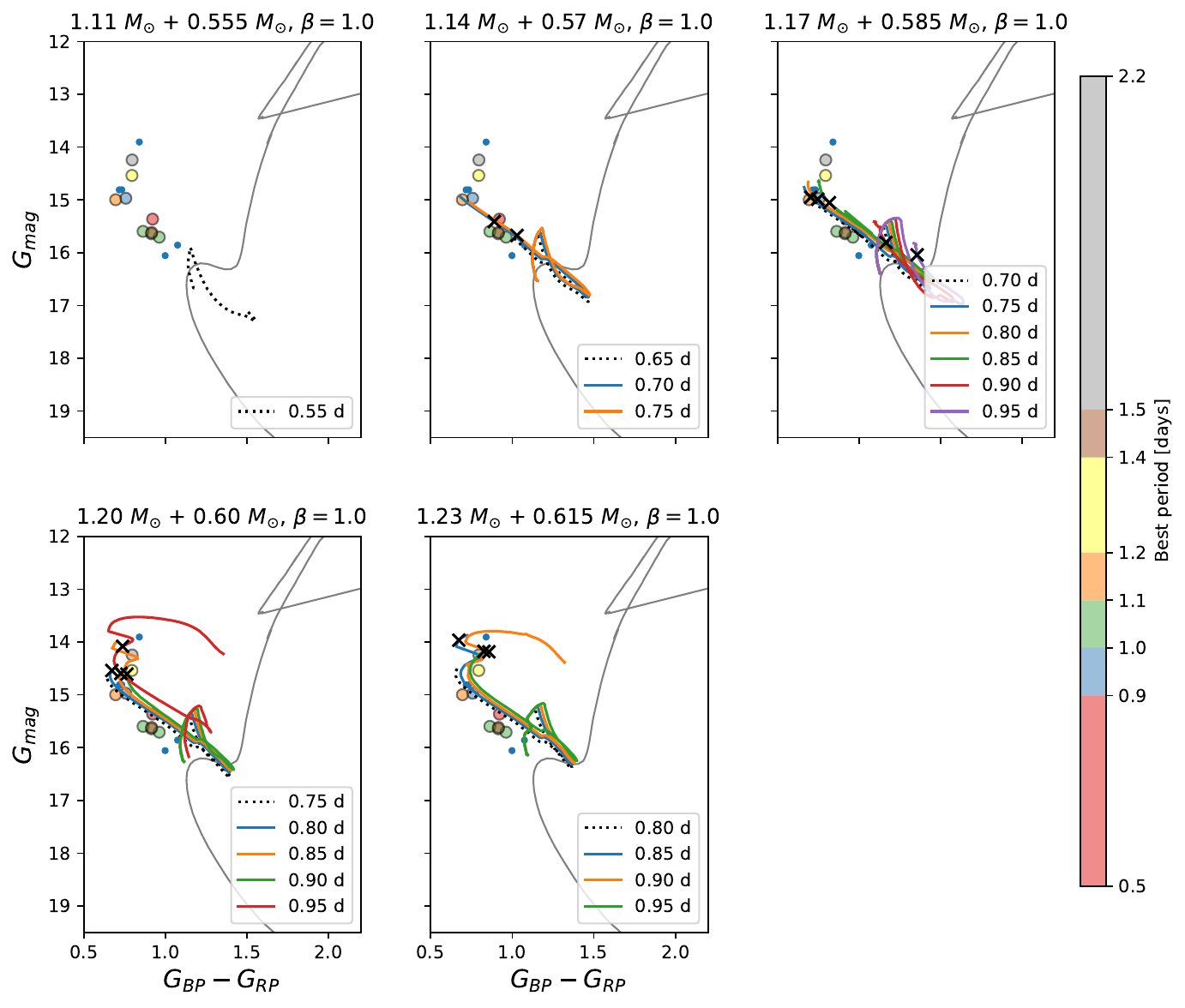}
    \caption{Same as Figure \ref{fig:q150beta100} but for the BSS progenitors listed in Table \ref{q200beta100}.}
    \label{fig:q200beta100}
\end{figure*}

In order to reach the fainter BSS in the CMD (i.e., those closer to the cluster turn-off), we performed a second set of calculations with a fixed mass ratio of $q = 2$, lowering the total mass of the system with respect to the previous set. The initial primary mass and orbital period were explored with the same steps as before. Table~\ref{q200beta100} lists the properties of the models in this second set, in the same format as Table~\ref{q150beta100}. As seen in Figure \ref{fig:q200beta100}, the overall behavior is similar to the previous set: most tracks reach $T_C$ in the brighter region of the CMD, with the exception of those systems whose accretor gained less mass due to their initial configuration. However, the lower total mass of this set allows some models to reach fainter magnitudes that were not reached with $q = 1.5$.

\begin{figure*}
    \centering
    \includegraphics[width=\linewidth]{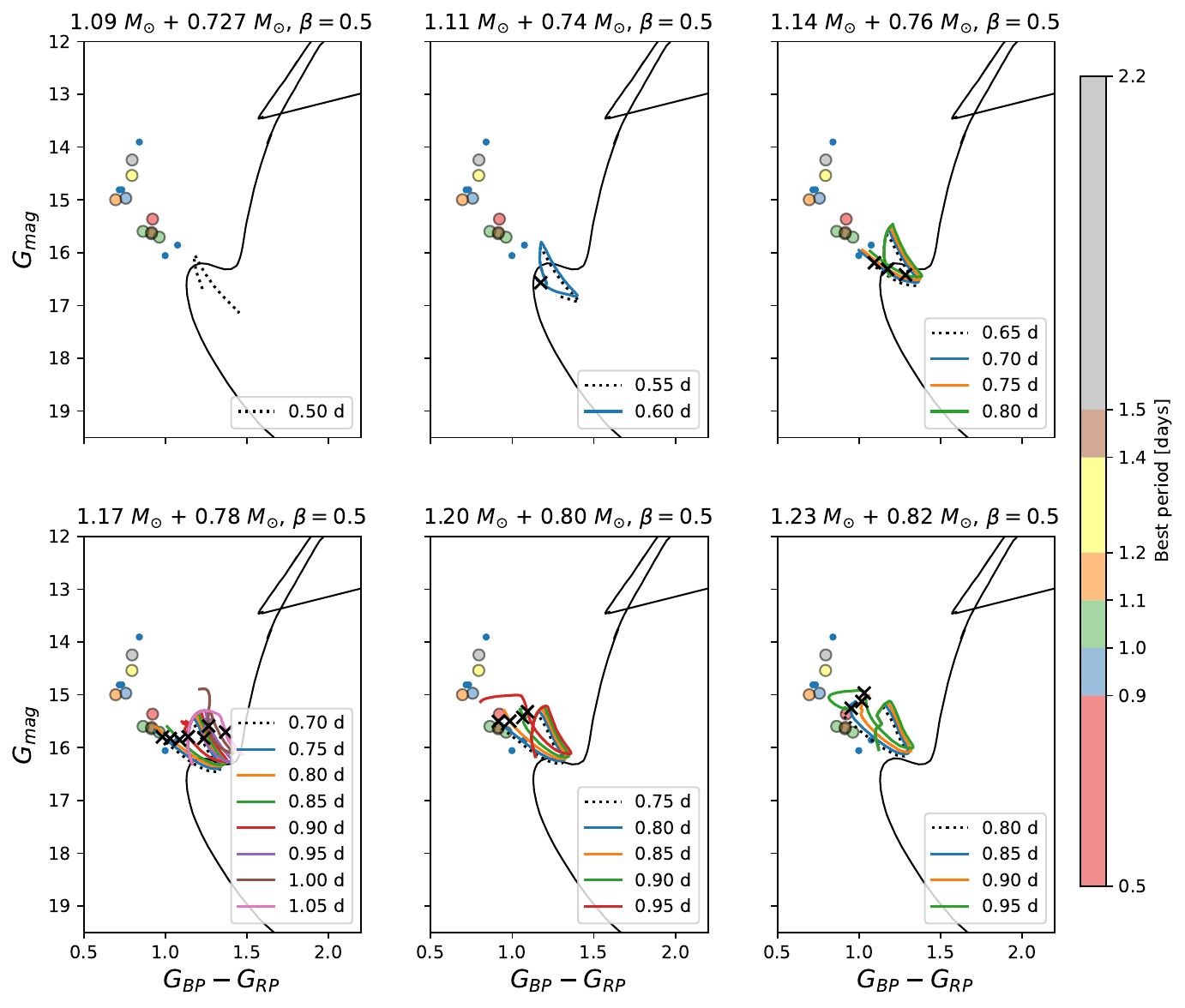}
    \caption{Same as Figure \ref{fig:q150beta100} and \ref{fig:q200beta100} but for the BSS progenitors listed in Table \ref{q150beta050}.}
    \label{fig:q150beta050}
\end{figure*}

We now explore a different approach to reproduce the fainter BSSs in our sample. Instead of assuming conservative MT, we calculate a third set of simulations using an intermediate MT efficiency $\beta=0.5$ and a mass ratio $q=1.5$. The initial primary mass and orbital period were explored using the same steps as in the other sets. As in Tables \ref{q150beta100} and \ref{q200beta100}, Table \ref{q150beta050} shows the properties of the models in this set. As seen in Figure \ref{fig:q150beta050}, all of these models reach the age $T_C$ at fainter magnitudes, i.e., greater than 15. The lightest binaries, i.e., those with initial primary masses of 1.09, 1.11, and 1.14, occupy positions around the TO or even below. In contrast, the models with higher initial mass easily account for the position of the BSSs with magnitudes $>15$ regardless of the value of $P_{\rm orb,i}$. Therefore, the mass lost from the binary during the MT process naturally limits the mass accreted onto the secondary, allowing the combined sequence to reach the position of the fainter BSSs in the CMD without the need to fine-tune the initial orbital period as in the conservative MT case.

\begin{table}
    \centering
    \begin{tabular}{c c c c c} 
     $M_1$  & $M_2$ & $P_{\rm orb,i}$ & STATUS & $P_{\rm orb,T_C}$ \\
     $[M_{\odot}]$ & $[M_{\odot}]$ & $[\rm days]$ &  & $[\rm days]$ \\
    \hline
   1.09  & 0.727& 0.55 & no MT        &      \\
   ''    & ''   & 0.50 & contact 6.13 & 0.31 \\
   1.11  & 0.74 & 0.65 & no MT        &      \\
   ''    & ''   & 0.60 & BSS          & 0.35 \\ 
   ''    & ''   & 0.55 & contact 6.25 & 0.31 \\
   1.14  & 0.76 & 0.85 & no MT        &      \\
   ''    & ''   & 0.80 & BSS          & 0.57 \\
   ''    & ''   & 0.75 & BSS          & 0.50 \\
   ''    & ''   & 0.70 & BSS          & 0.41 \\
   ''    & ''   & 0.65 & contact 6.60 & 0.32 \\
   1.17  & 0.78 & 1.10 & no convergence  \\
   ''    & ''   & 1.05 & BSS          & 1.81 \\
   ''    & ''   & 1.00 & BSS          & 1.52 \\
   ''    & ''   & 0.95 & BSS          & 1.26 \\
   ''    & ''   & 0.90 & BSS          & 0.94 \\
   ''    & ''   & 0.85 & BSS          & 0.70 \\
   ''    & ''   & 0.80 & BSS          & 0.52 \\
   ''    & ''   & 0.75 & BSS          & 0.39 \\   
   ''    & ''   & 0.70 & contact 6.41 & 0.33 \\
   1.20  & 0.80 & 1.00 & BSS          & 2.32 \\
   ''    & ''   & 0.95 & BSS          & 1.47 \\
   ''    & ''   & 0.90 & BSS          & 1.30 \\
   ''    & ''   & 0.85 & BSS          & 0.68 \\
   ''    & ''   & 0.80 & BSS          & 0.40 \\
   ''    & ''   & 0.75 & contact 6.37 & 0.33 \\
   1.23  & 0.82 & 1.00 & BSS          & 2.36 \\
   ''    & ''   & 0.95 & BSS          & 1.66 \\
   ''    & ''   & 0.90 & BSS          & 1.49 \\
   ''    & ''   & 0.85 & BSS          & 0.60 \\
   ''    & ''   & 0.80 & contact 6.48 & 0.36 \\
   
    \hline
    \end{tabular}
    \caption{Same as Table \ref{q150beta100} and \ref{q200beta100} but for those models in the third set of simulations ($q=1.5$, $\beta=0.5$).}
    \label{q150beta050}
\end{table}

\section{Summary and final remarks}
\label{sec4}

Numerous observations of BSSs in diverse stellar systems support the hypothesis that these objects form through different channels (\citealt{Piotto_2004}). While direct collisions are considered an important mechanism to produce BSSs in dense stellar environments, MT in interacting binaries is believed to be the easiest way to form BSSs in sparse environments as open clusters (e.g. \citealt{Sollima_2008}, \citealt{Rain_2024}). Here, we explore the MT mechanism in forming the short-period BSSs in the OC Collinder~261. To assess the stellar variability of BSS candidates, we utilized public data and extracted light curves from the PATHOS project, aiming to identify close binary systems among these candidates. After identifying BSSs with short orbital periods (shorter than 2.1~d), we designed a set of binary models to find possible progenitors of these BSSs. We retain those models that undergo an MT episode and reach the BSSs' region in the CMD by the age of the cluster while having orbital periods compatible with the observations. We then conclude that the short-period eclipsing BSSs in Collinder 261 could have been formed through MT. Moreover, the models obtained correspond to semidetached or contact binary systems. This is consistent with the classifications reported by \citet{Rain_2020} based on the SIMBAD astronomical database \citep{Wenger_2000} for seven of the short-period BSSs in our sample. Six of these are identified as $\beta$~Lyrae and Algol-type binaries (both representative of semidetached configurations), and one as a W-UMa-type star (where the components are in contact or almost in contact, and share a common envelope of material). These classifications further support the consistency between our models and the observed properties of these BSSs. Models that reach a contact configuration before the age of the cluster are natural progenitor candidates for the shortest-period BSSs in our sample ($P \lesssim 0.5$ days), which are the most likely to currently reside in a contact configuration. This interpretation is supported by the observed presence of W~UMa-type binaries within the BSS region of open and globular clusters \citep{Kaluzny1987ApJ...314..585K,Rucinski2000AJ....120..319R}, and is consistent with the known ability of low-mass contact binaries to survive without merging for timescales comparable to the cluster age \citep{Tutukov2004Evolutionary,Jiang2026Merger}. Interestingly, the five SX~Phoenicis pulsator candidates identified among the BSS variables also occupy the same region of the CMD as our contact binary models.. If these systems are the products of the coalescence of contact binaries, their positions in the CMD would be consistent with our simulations, since the merger product would appear as a single, more massive star. Whether such merger products would fall within the pulsational instability strip and be observable as SX~Phe pulsators is an interesting possibility that lies beyond the scope of this work, as our code does not follow the evolution past the onset of contact.

This study did not aim to find all possible progenitors for the observed BSSs, but rather to show that they can be formed via the MT mechanism. Our findings on the variation of the MT efficiency indicate that this parameter plays a significant role in the formation of these objects, which warrants further exploration. Historically, 1D binary models often assumed constant MT efficiency, although a better approach on this matter is critical and still lacking \citep{Chen2002,2004MNRAS.355.1182C,deMink2007}. In particular, whether the accretor can continue to gain mass after reaching critical rotation is still debated, and it has been proposed that accretion might still proceed if angular momentum can be redistributed through a disk \citep{Popham1991}. This mechanism could facilitate the formation of BSSs and may also be related to the recently identified population of “blue lurkers”, which appear to be products of similar processes but with lower final masses \citep{Leiner_2019}. Advancing our understanding of MT in binaries will, therefore, benefit from studies encompassing a wide range of binary evolutionary outcomes, including but not limited to BSSs.

The relevance of inefficient MT suggested by our models is consistent with the findings reported by \citet{2021ApJ...908....7S}, \citet{2023ApJ...944...89S} and \citet{Leiner_2021}. More recently, \citet{Wang2025} analyzed the BSS population in the $\sim$1.58 Gyr-old NGC 2173 globular cluster using a population synthesis approach. Their results show that inefficient MT plays a crucial role in the formation of the blue-sequence BSSs, objects traditionally associated with direct stellar collisions. Taken together, these findings highlight that the efficiency of MT is a key factor shaping the properties and distribution of BSSs, not only in sparse stellar environments but also within denser regions.

We emphasize that this work focuses exclusively on the MT channel in isolated primordial binaries, without invoking dynamical interactions with other cluster members. While stellar collisions and dynamically-driven mergers are relevant in dense environments such as globular cluster cores, the low stellar density of open clusters like Collinder~261 makes isolated binary evolution the most natural formation channel for their BSS populations \citep{Sollima_2008, Rain_2024}. To our knowledge, this is the first study to simultaneously constrain both the CMD position and the orbital period of a large sample of short-period BSSs through detailed binary evolution models, including the calculation of combined atmospheric sequences that account for the unresolved flux of both components as observed photometrically.

\begin{acknowledgements}
\\
We gratefully acknowledge the support of the {\it Coimbra Group Scholarship Program for Young Professors and Researchers from Latin American Universities}, which facilitated M. Echeveste’s stay at the University of Padova.  This opportunity was instrumental in the early stages of this project.  
\\
We thank the valuable comments by Dr. E. Pancino, which have helped us improve this article.\\
Co-funded by the European Union (ERC-2022-AdG, {\em "StarDance: the non-canonical evolution of stars in clusters"}, Grant Agreement 101093572, PI: E. Pancino). Views and opinions expressed are, however, those of the author(s) only and do not necessarily reflect those of the European Union or the European Research Council. Neither the European Union nor the granting authority can be held responsible for them.\\
OGB is a Member of the Carrera del Investigador Cient\'{\i}fico of the Comisi\'on de Investigaciones Cient\'{\i}ficas de la Provincia de Buenos Aires (CIC PBA), Argentina.
\end{acknowledgements}

\bibliography{biblio}{}
\bibliographystyle{aasjournalv7}



\begin{appendix}
\label{app:lightcurves}

\begin{figure*}
    \centering
    \includegraphics[width=\linewidth]{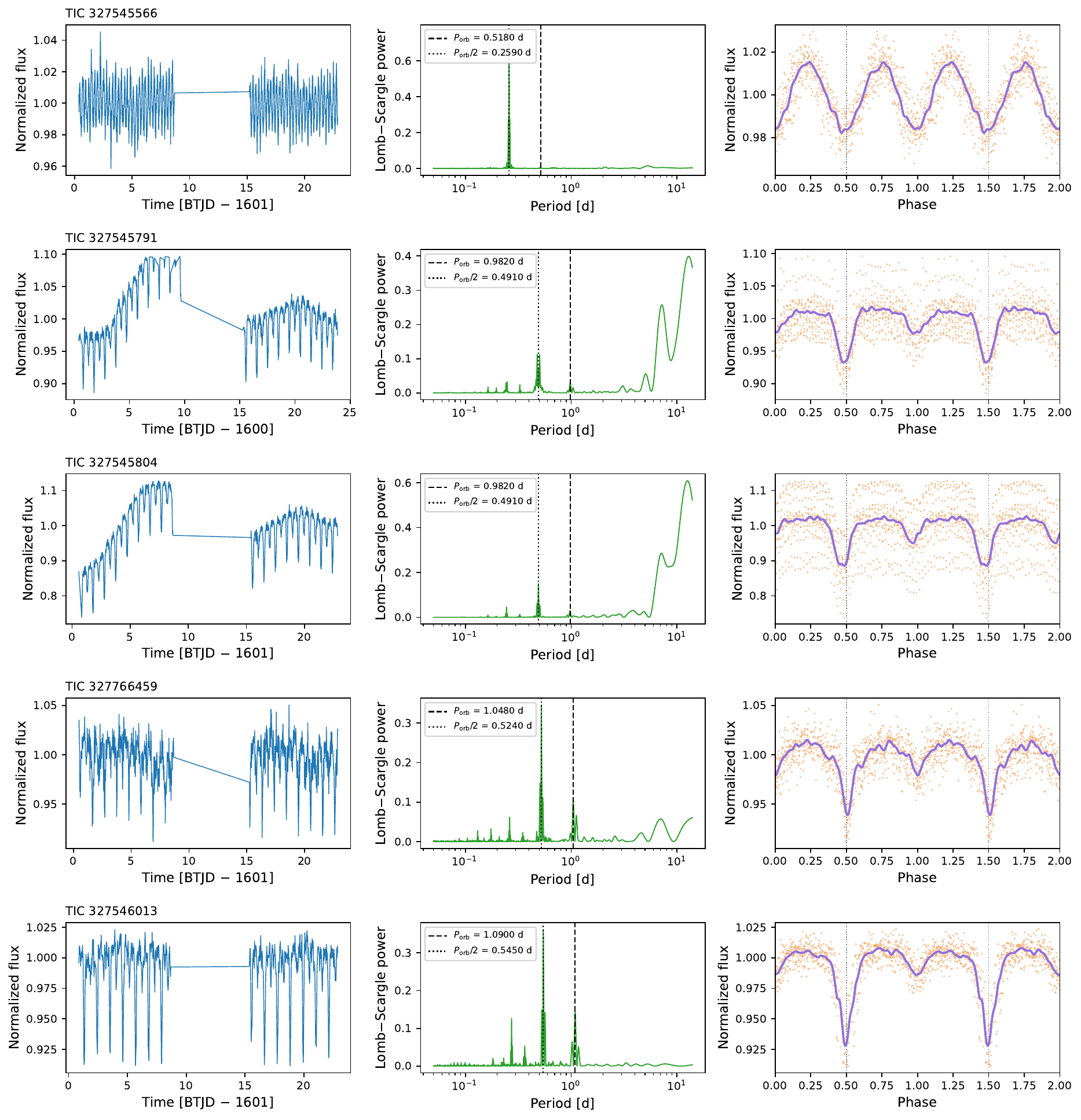}
    \caption{TESS/PATHOS light curves of the eclipsing/ellipsoidal binary BSS candidates in Collinder\,261, from Sector\,11. Each row corresponds to one star, ordered by increasing orbital period. Left: normalized PATHOS PSF light curve. Middle: Lomb--Scargle periodogram; the dashed line marks the adopted orbital period $P_{\rm orb} = 2\,P_{\rm LS}$ and the dotted line marks the dominant peak at its first harmonic $P_{\rm orb}/2$, corresponding to the two minima per orbit. Right: light curve phase-folded on $P_{\rm orb}$, with the smoothed binned median overplotted (purple); the primary minimum is placed at phase 0.5. All objects are listed in Table~\ref{q200beta100}.}
    \label{fig:app_eclipsing1}
\end{figure*}

\begin{figure*}
\centering
\includegraphics[width=\linewidth]{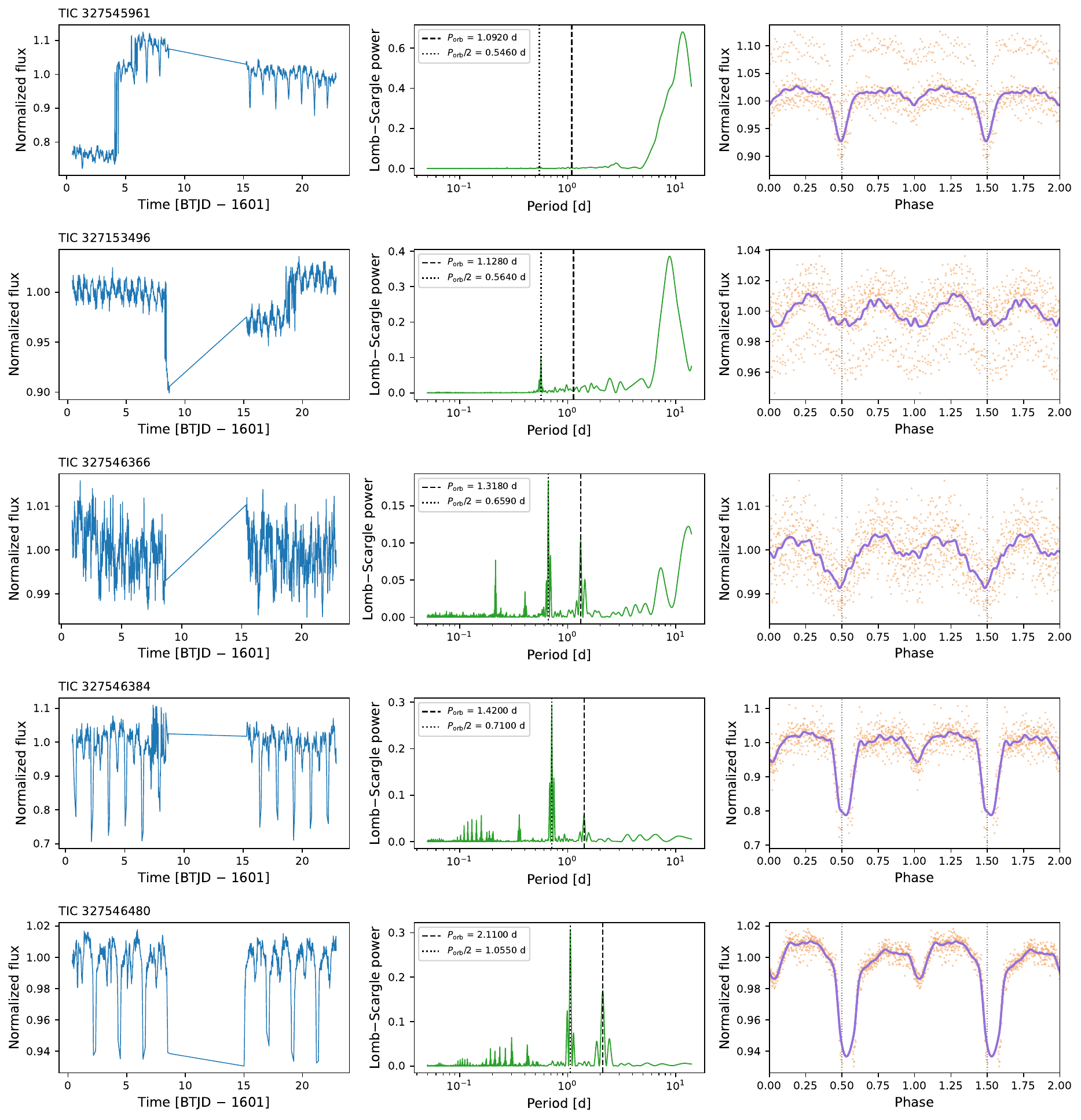}
\caption{Continuation of Figure~\ref{fig:app_eclipsing1}: remaining
eclipsing/ellipsoidal binary BSS candidates.}
\label{fig:app_eclipsing2}
\end{figure*}

\begin{figure*}
\centering
\includegraphics[width=\linewidth]{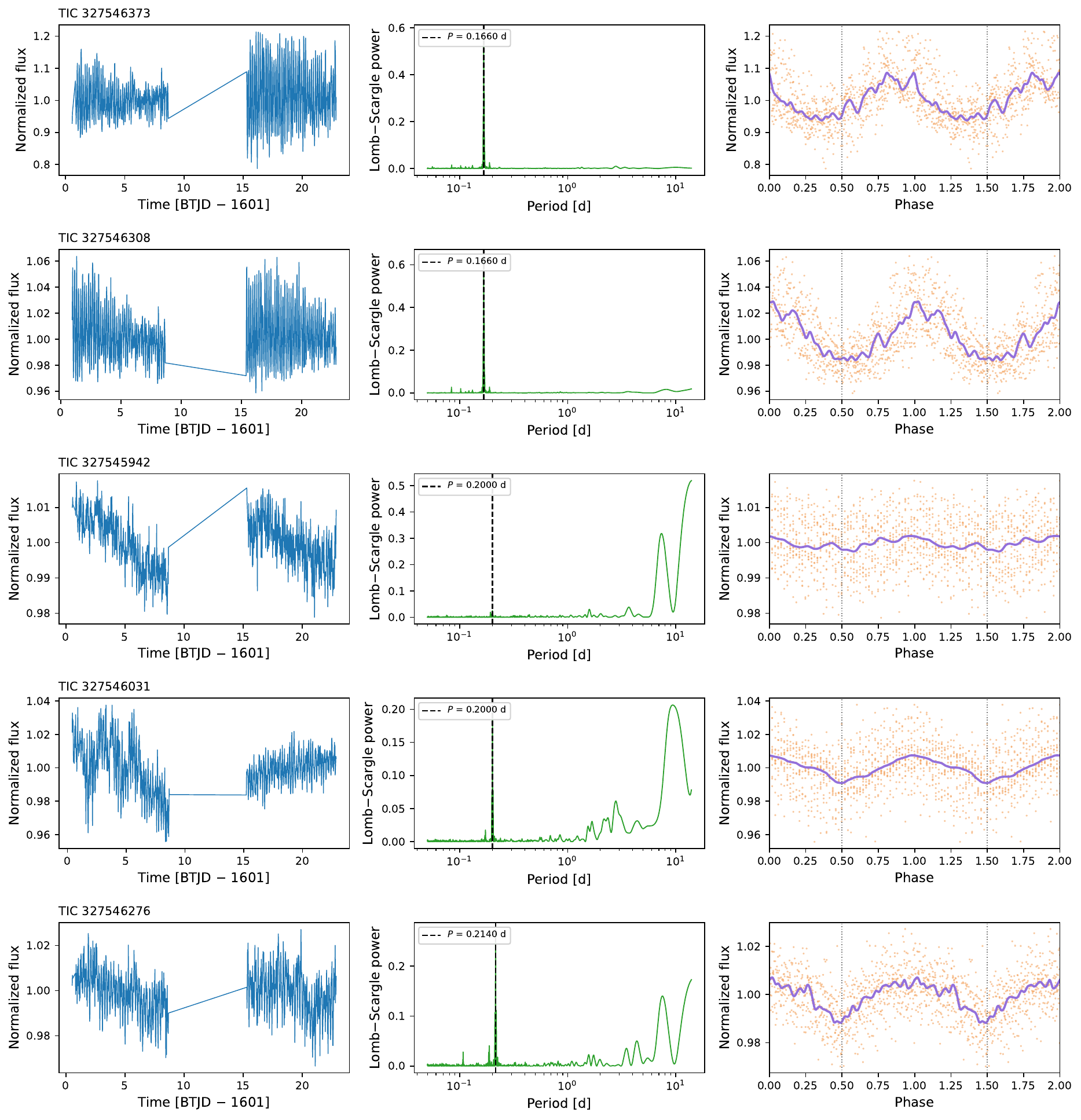}
\caption{Same as Figure~\ref{fig:app_eclipsing1}, but for the SX~Phoenicis
pulsator candidates. For these stars the light curve shows a single minimum
per cycle, so the pulsation period equals the dominant Lomb--Scargle peak
($P_{\rm puls} = P_{\rm LS}$, dashed line) and the folded curves use
$P_{\rm puls}$ directly. Continued in Figure~\ref{fig:app_sxphe2}.}
\label{fig:app_sxphe1}
\end{figure*}

\begin{figure*}
\centering
\includegraphics[width=\linewidth]{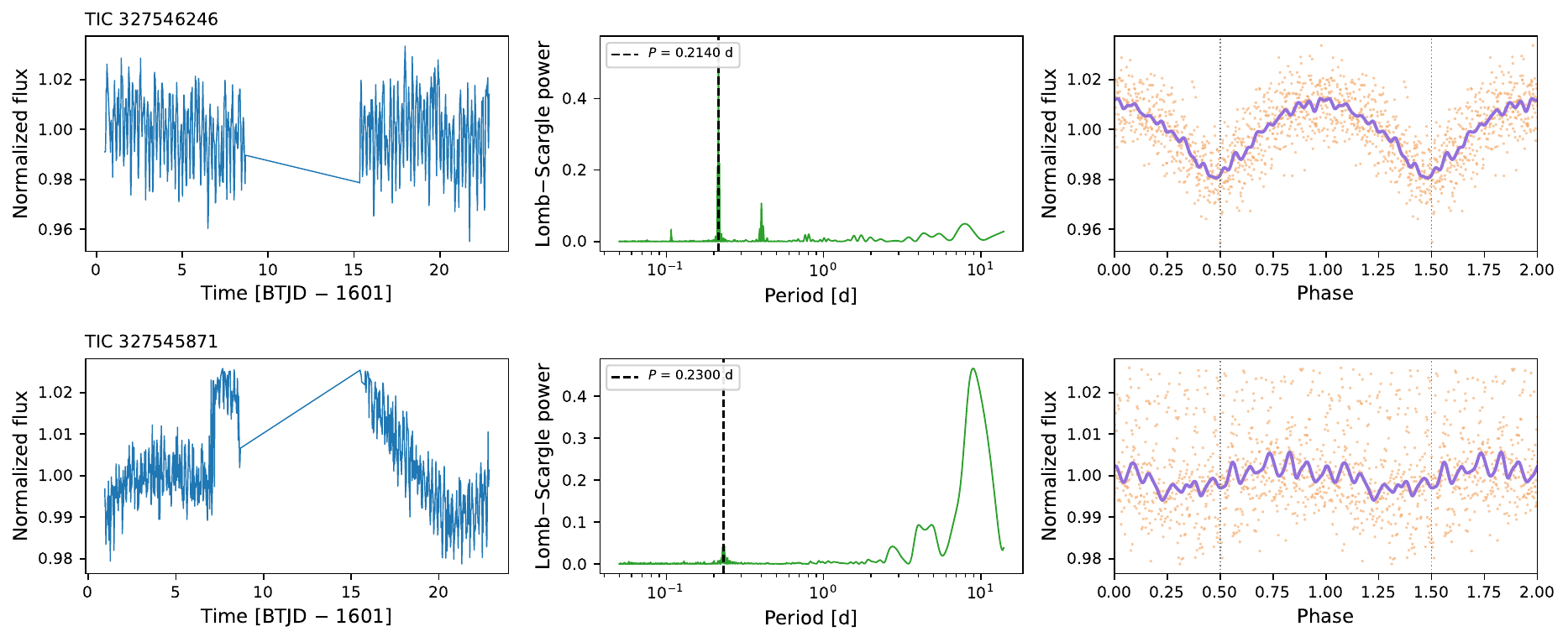}
\caption{Continuation of Figure~\ref{fig:app_sxphe1}: remaining
SX~Phoenicis pulsator candidates.}
\label{fig:app_sxphe2}
\end{figure*}

\begin{figure*}
\centering
\includegraphics[width=\linewidth]{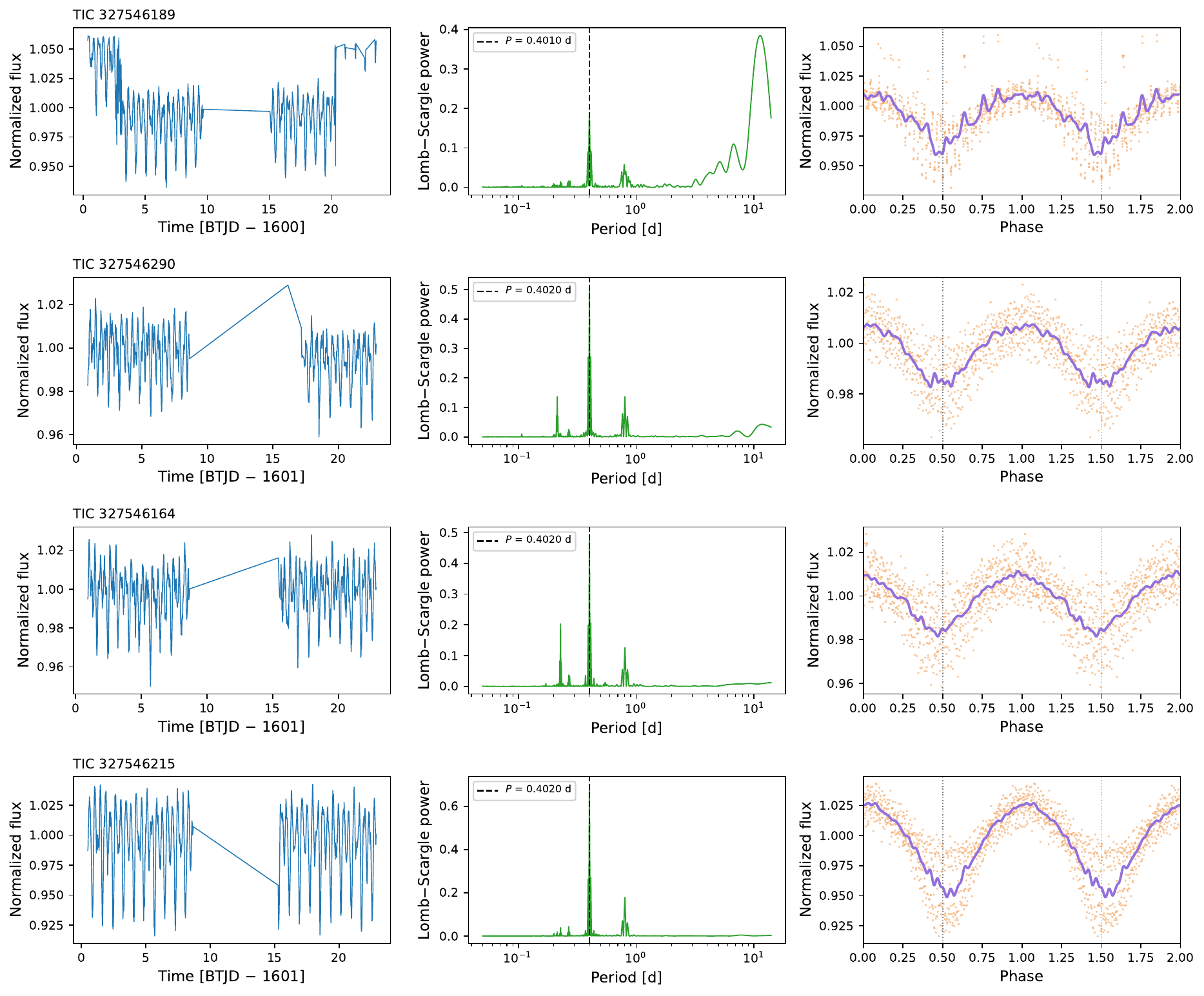}
\caption{Same as Figure~\ref{fig:app_eclipsing1}, but for the four stars whose dominant periodicity ($P \approx 0.402$~d, dashed line) is shared among close neighbors on the detector, and for the three stars whose signal is a blend of a neighbouring variable BSS (TIC~327545791, TIC~327546276 and TIC~327546373).}
\label{fig:app_alias}
\end{figure*}
\end{appendix}

\end{document}